\documentclass{article}

\usepackage[english]{babel}

\usepackage[letterpaper,top=2cm,bottom=2cm,left=3cm,right=3cm,marginparwidth=1.75cm]{geometry}

\usepackage{amsmath}
\usepackage{comment}
\usepackage{soul}
\usepackage[colorlinks=true, allcolors=blue]{hyperref}

\usepackage{booktabs}
\usepackage{tabularx}
\usepackage{makecell}

\usepackage{siunitx}
\usepackage{caption}
\usepackage{array}
\usepackage{longtable}
\usepackage{geometry}

\usepackage{graphicx}   
\usepackage{authblk}
\usepackage{cite}
\usepackage{subcaption}
\usepackage{comment}
\usepackage[colorinlistoftodos]{todonotes}

\usepackage{adjustbox}
\usepackage[utf8]{inputenc}
\usepackage[T1]{fontenc}

\AtBeginDocument{%
  \newgeometry{top=2cm,bottom=2cm,left=2cm,right=2cm}%
}

\makeatletter
\newcommand*\samethanks[1][\value{footnote}]{\footnotemark[#1]}
\makeatother

\title{Suspended thin-film lithium niobate modulator for broadband mid-infrared light modulation and frequency comb generation}

\author[1,2]{Chun-Ho Lee\thanks{These authors contributed equally.}}
\author[1,2]{Xinyi Ren\samethanks}
\author[2]{Xinzhou Su\samethanks}
\author[3]{Wonho Lee}
\author[2]{Zile Jiang}
\author[1,2,6]{Yue Yu}
\author[2]{Huibin Zhou}
\author[2]{Yue Zuo}
\author[2]{Shaoyuan Ou}
\author[1,2]{Reshma Kopparapu}
\author[4]{Adam T. Heiniger}
\author[5]{Moshe Tur}
\author[2]{Alan E. Willner}
\author[1,2]{Zaijun Chen}
\author[1,2,6]{Mengjie Yu\thanks{Corresponding author: mengjie.yu@berkeley.edu}}

\affil[1]{Department of Electrical Engineering and Computer Sciences, University of California, Berkeley, CA, USA}
\affil[2]{Ming Hsieh Department of Electrical and Computer Engineering, University of Southern California, Los Angeles, CA, USA}
\affil[3]{PHY research lab, Intel lab, Hillsboro, OR, USA}
\affil[4]{TOPTICA Photonics Inc., Pittsford, NY 14534, USA}
\affil[5]{School of Electrical Engineering, Tel Aviv University, Ramat Aviv 69978, ISRAEL}
\affil[6]{Materials Sciences Division, Lawrence Berkeley National Laboratory, Berkeley, California 94720, USA}

\date{} 

\begin{document}
\maketitle

\begin{abstract}

The mid-infrared (MIR) spectral regime is central to applications including remote sensing, precision spectroscopy, higher harmonic generation, and free-space optical communication. However, coherent and broadband MIR modulation remains challenging owing to high optical loss, limited bandwidth, and large drive voltages in existing platforms.  Here, we overcome the challenges by deploying a suspended thin-film lithium-niobate (TFLN) based electro-optic (EO) platform co-designed with high-performance traveling-wave microwave (MW) electrodes. We demonstrate a record-low $V_{\pi,DC}$ of 2.3 to 4.3 V over a broadband MIR bandwidth from 2.4 to 3.6 $\mu$m, and a 2.7-dB EO bandwidth of 40 GHz (extracted 3-dB bandwidth of 50 GHz), yielding a figure-of-merit of 17.4 GHz/V—more than an order of magnitude higher than the state-of-the-art. We demonstrate, for the first time, high frequency $V_{\pi,MW}$ of  4.5-6.5 V in the 25-35 GHz range, and frequency-agile MIR EO frequency comb generation with a 10-dB optical bandwidth over 0.8 THz using a suspended phase modulator of 4-cm active modulation length. We further validate the platform in a free-space optical communication link. Our results establish a monolithic MIR photonic platform capable of powerful EO modulation and spectral synthesis, and present a significant step towards reconfigurable MIR sensing and communication systems on chip.

\end{abstract}

\section{Main text}
 Rapidly growing amount of data being transmitted across networks is creating a significant demand for new wireless communication technologies to support higher data rate, lower latency, higher connection density, and global coverage\cite{kazakov2025}. Free space optical communication is an emerging technology for data transfer to remote assets through atmospheric communication channels, addressing the broadband connectivity bottlenecks for space and terrestrial applications \cite{khatri2025,zou2022} while offering tremendous advantages over radio-frequency carriers due to its high bandwidth, immunity to electromagnetic interferences, low latency, and multiplexing features.
 
 The MIR spectral range from 3 to 5 $\mu$m and long-wave infrared (LIR) from 8 to 12 $\mu$m are actively being investigated as free space optical carriers since it has the lower absorption in atmosphere, the higher tolerance to adverse weather conditions (such as dust, haze, and low-altitude clouds), and less phase-front distortion by turbulence effects compared to the near-infrared (NIR), mm-wave, and THz-waves \cite{mcclatchey1979, petersen2014,goldstein2022,kamboj2022,chiles2014,yan2017,mishra2021,hwang2023,ledezma2023,ko2025,nader2019}. There are recent developments of the optoelectronic devices based on subsequent intersubband transitions and stark effect at LIR, such as quantum cascade laser (QCL), quantum-well infrared detector (QWIP) and quantum cascade detector, however the efficiency of unipolar device is significantly worse in the MIR region\cite{hoghooghi2024,soibel2010,dougakiuchi2021,dely2022,pirotta2021}. Existing approaches to modulate the MIR light include direct current modulation of a QCL at a limited bandwidth of a few GHz at room temperature\cite{hinkov2016,pang2020,lotfi2016,soibel2010,pang2022,benirschke2020}, and nonlinear parametric conversion of the near-infrared light which suffers from limited optical conversion bandwidth and efficiency as well as requires external power-hungry pump lasers\cite{willner2024}. Alternatively, EO material could be used to modulate the refractive index or optical absorption via external voltages. The MIR EO modulators   have been demonstrated in silicon \cite{nedeljkovic2014}, silicon on lithium niobate (LN) \cite{xu2023,chiles2014}, titanium dioxide on LN \cite{jin2019}, ion diffused waveguide on LN \cite{becker1985}, black phosphorous on silicon \cite{huang2021,nedeljkovic2019}, germanium on silicon \cite{malik2014,li2019}, and barium titanate \cite{jin2019BTO}. However, none of the material platform offers beyond a few GHz modulation speed and CMOS-compatible half-wave voltages. In addition, traditional silicon-based modulators have inevitable high optical losses due to the free carrier absorption at longer optical wavelength. Therefore, a compact, high-speed, efficient and low-loss EO modulator in the MIR is still missing.

\begin{figure}[h]
\centering
\includegraphics[scale=1]{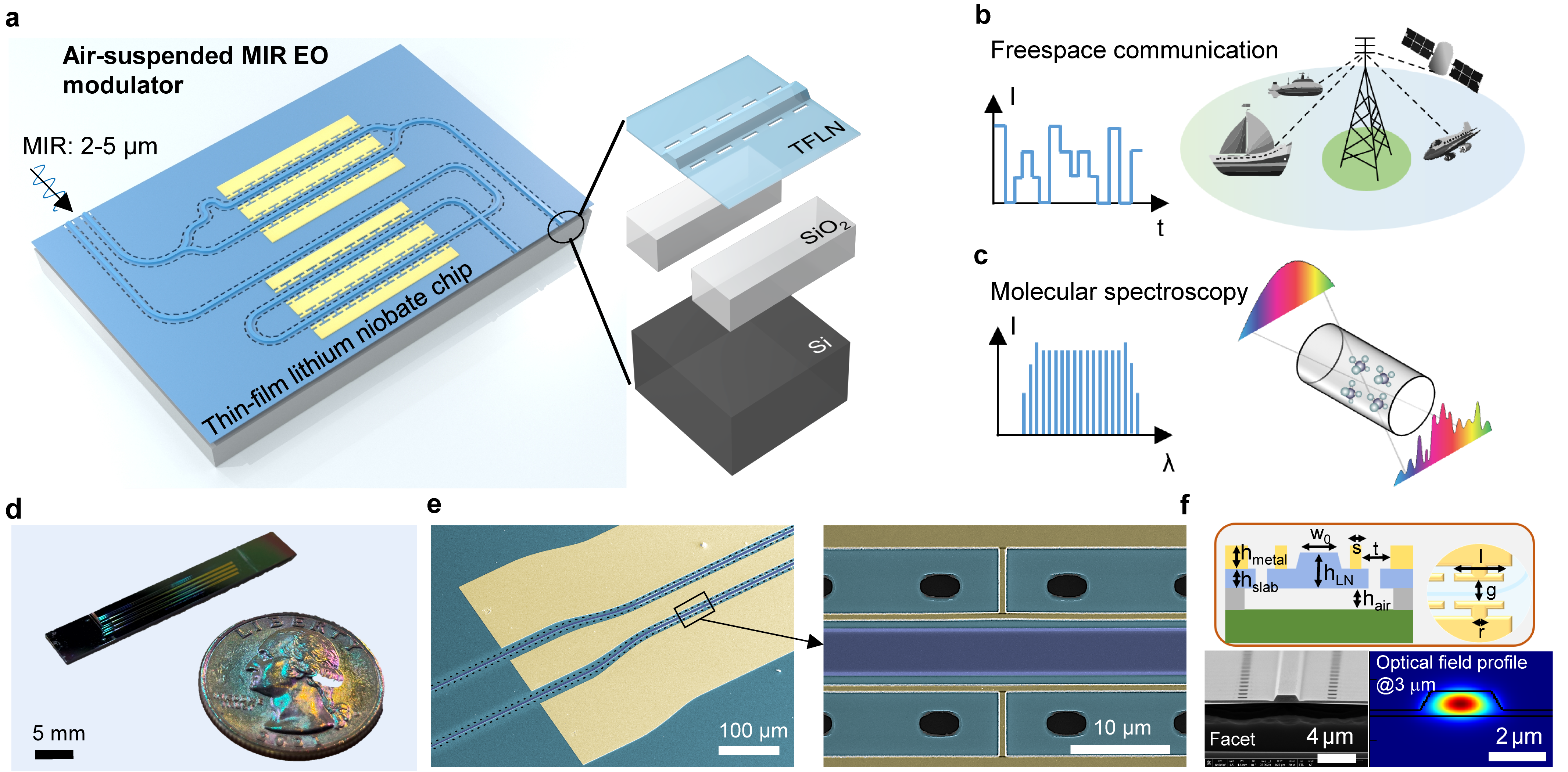}
\caption{\label{fig:figure1} \textbf{Monolithic mid-infrared optoelectronic platform on TFLN.} (a) Air-suspended MIR traveling-wave-based amplitude and phase electro-optic modulators on the TFLN. Inset shows the cross-sectional view of the optical layer which consists of a ridge TFLN waveguide sitting on oxide-on-Si substrate. The oxide layer is removed underneath the waveguide area through air holes on the TFLN slab layer etched via a second etching process. High performance optoelectronic interface in the MIR would enable high-bandwidth data links via free space communication (b) and frequency-agile EO frequency combs for molecular spectroscopy (c). (d) Photographic image of the fabricated MIR modulator chip on 800-nm X-cut TFLN. (e) Scanning-electron-microscopy (SEM) image of suspended photonic waveguides and coplanar microwave transmission line. Segmented slow-wave electrode design (zoom in, right) is applied to reduce the microwave loss and achieve impedance matching and velocity matching with optical field while compatible with a low optical propagation loss. Air hole arrays are optimized and placed between the metal segments for releasing the adjacent photonic waveguides. (f) Cross section of the suspended device where the metal gap (g), waveguide height ($h_{LN}$), slab thickness ($h_{slab}$), metal thickness ($h_{metal}$), waveguide width ($w_{0}$), the air gap under the waveguide ($h_{air}$) and the segmented electrode parameters (s , t, r , l ) are 6.5, 0.8, 0.25, 0.8, 4, 4.7 $\mu$m, and (0.5, 6.5, 0.5, 45) $\mu$m, respectively. The SEM image of the waveguide facet is shown. The simulated transverse-electric optical mode profile at 3 $\mu$m is plotted.}
\end{figure}

 Here, we present a monolithic optoelectronic platform in the MIR based on a suspended TFLN platform, capable of both high-speed amplitude and phase modulation (Fig.~\ref{fig:figure1}a). The air suspended TFLN is used to overcome the limitation of the absorption loss in the MIR from the underlying silicon dioxide layer as well as to support a tight optical confinement in the waveguide and reduce the MW propagation loss, both of which are critical to achieve a low switching voltage \cite{mishra2021}. Co-designed with the optical waveguide, the segmented coplanar MW electrodes are deployed to further reduce the ohmic loss and achieve velocity matching between the MW and optical signals, supporting high speed EO response. The demonstrated traveling-wave-based Mach-Zehnder amplitude modulator (AM) and double-pass phase modulator (PM) are based on (non-resonant) waveguide structures which can operate in a broad MIR wavelength range and be compatible with integrated MIR light sources. Unlike electro-absorption or carriers based modulators \cite{malik2014,li2019}, our modulator platform based on the Pockel’s effect not only allows for high-fidelity data links for free space optical communication via independent amplitude and phase modulation, but also can be used to generate broadband frequency-agile optical frequency comb  \cite{yu2022} for spectroscopy and sensing applications (Fig.~\ref{fig:figure1}b\&c) \cite{hoghooghi2024}. Until now, none of the existing platforms have achieved broadband tunable MIR EO frequency comb generation \cite{hwang2023,didier2025,hwang2025}.

 Figure~\ref{fig:figure1}d presents an optical microscope image of the integrated MIR modulator chip with a footprint of $25\times 7~\mathrm{mm}^2$. The chip is fabricated on an 800-nm X-cut LN on 4.7-$\mu$m oxide on Si substrate wafer. The waveguide has a top width of 4 $\mu$m and an etch depth of 500 nm, patterned via electron-beam lithography and ion milling (see Methods). Air holes are then defined and etched on the remaining 300-nm LN slab via a second step of lithography and etching, followed by the metal-electrode patterning, deposition and lift-off process. The  air holes are placed in between the metal segments and used to selectively release the bottom oxide only around the photonic waveguide via a wet etching process, providing both good mechanical and thermal support for the electrodes. Scanning electron microscopy (SEM) image in Fig.~\ref{fig:figure1}e shows the structures of the air holes, suspended optical waveguide, and the segmented microwave electrodes. We optimize the operation for a fundamental transverse electric (TE) mode. Here, the smallest electrode gap of 6.5 $\mu$m along with the waveguide width of 4 $\mu$m is chosen to maximize the EO response as well as to minimize the metal induced optical loss. The cross-section of the EO waveguide and microwave transmission line are illustrated in Fig.~\ref{fig:figure1}f along with the optimized device geometry parameters after co-design and the simulated optical mode profile at 3 $\mu$m. The chip is cleaved for edge coupling at both input and output facets.  

\begin{figure}[h]
\centering
\includegraphics[scale=1]{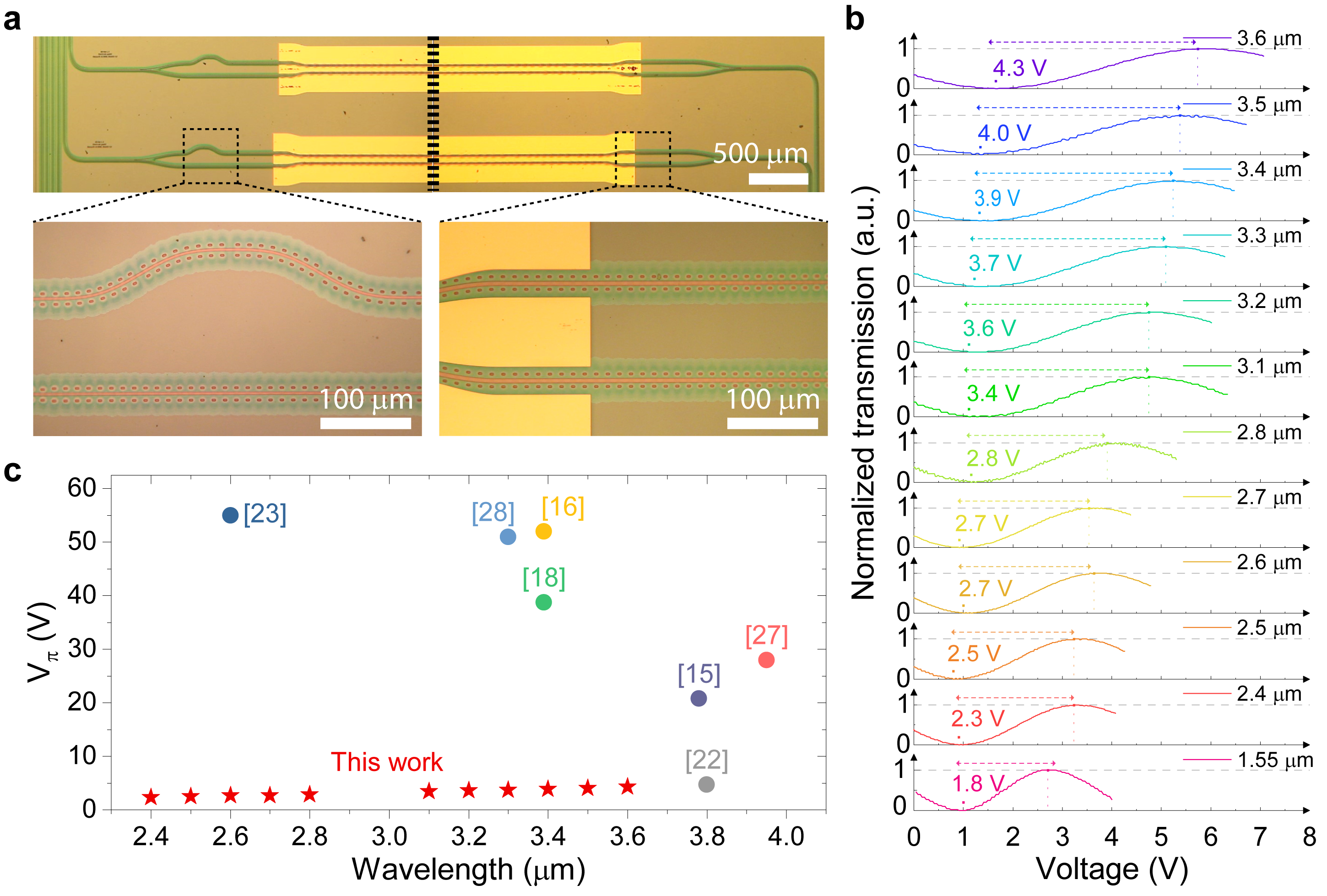}
\caption{\label{fig:figure2} \textbf{Characterization of near-DC half-wave voltage $V_\pi$ of a 2-cm-long air-suspended amplitude modulator.} (a) Optical microscope images. Two optical path in the AM have a different length of 21 $\mu$m to enable bias point tuning via varying the optical wavelength. (b) Normalized optical transmission as a function of applied voltage at the different MIR wavelengths as well as at the 1.55 $\mu$m in the NIR. The measured half-wave voltage ($V_\pi$) at 100 kHz is 2.3 - 4.26 V across the optical span from 2.4 $\mu$m to 3.6 $\mu$m, which indicates approximately linear dependence of the optical operational wavelength. In addition, the suspended AM device is measured across more than an octave optical bandwidth with a half-wave voltage of 1.8 V at 1.55 $\mu$m. (c) Comparison of  $V_\pi$ with other MIR EO platforms. The LN amplitude modulator shows the lowest  $V_\pi$  as well as  $V_\pi\cdot$L of 4.6~V$\cdot$cm at 2.4~$\mu$m and 8.6~V$\cdot$cm at 3.6 $\mu$m.}
\end{figure}

We first characterize the near-DC half-wave voltage, namely $V_\pi$, of our AM device at a low modulation frequency of 100 kHz. The suspended AM shown in Fig.~\ref{fig:figure2}a includes two Y-shape splitters, two asymmetric arms of a 21-$\mu$m length difference and an active modulation length of 2 cm in a push-pull configuration. The continuous-wave tunable optical parametric oscillator (Toptica, TOPO) is used to inject the device via a free space objective lens at a MIR wavelength range from 2.4 to 3.6 $\mu$m. We record the modulated optical output using an AC-coupled mercury cadmium telluride (MCT) detector while applying a saw-tooth voltage waveform to the electrodes at 100 kHz. Figure~\ref{fig:figure2}b  plots the optical transmission as a function of applied voltage at various MIR wavelengths, featuring the $V_\pi$ of 2.3 V at 2.4 $\mu$m to 4.3 V at 3.6 $\mu$m. The corresponding half-wave voltage length product ($V_\pi \cdot L$) are 4.6 $V\cdot cm$ at 2.4 $\mu$m and 8.6 $V\cdot cm$ at 3.6 $\mu$m wavelength, respectively. The $V_\pi$  scales near linearly with the optical wavelength $\lambda$, considering that the accumulated phase shift $\Delta \phi \propto \Gamma / \lambda$ where  $\Gamma$  is the mode overlap (see Methods). As the suspended optical waveguide supports guided mode over an ultrabroad bandwidth, we tested the $V_\pi$ at the NIR wavelength of 1.55 $\mu$m to be 1.8 V, demonstrating our integrated EO modulator operation across a 1.2-octave span. The extinction ratio is measured to be 7 dB in the MIR using a lock-in amplifier and can be improved by deploying a single mode waveguide at the Y splitter to suppress the scattering into  undesired optical modes. We compare the near-DC $V_\pi$ with the state-of-art reported in other MIR modulators at similar wavelength range in Fig.~\ref{fig:figure2}c. This work  presents the lowest $V_\pi$ ever reported in the MIR, reaching CMOS-compatible voltages, and the broadest MIR optical bandwidth, both of which are critical for in-parallel power-efficient modulation.

\begin{figure}[h]
\centering
\includegraphics[scale=1]{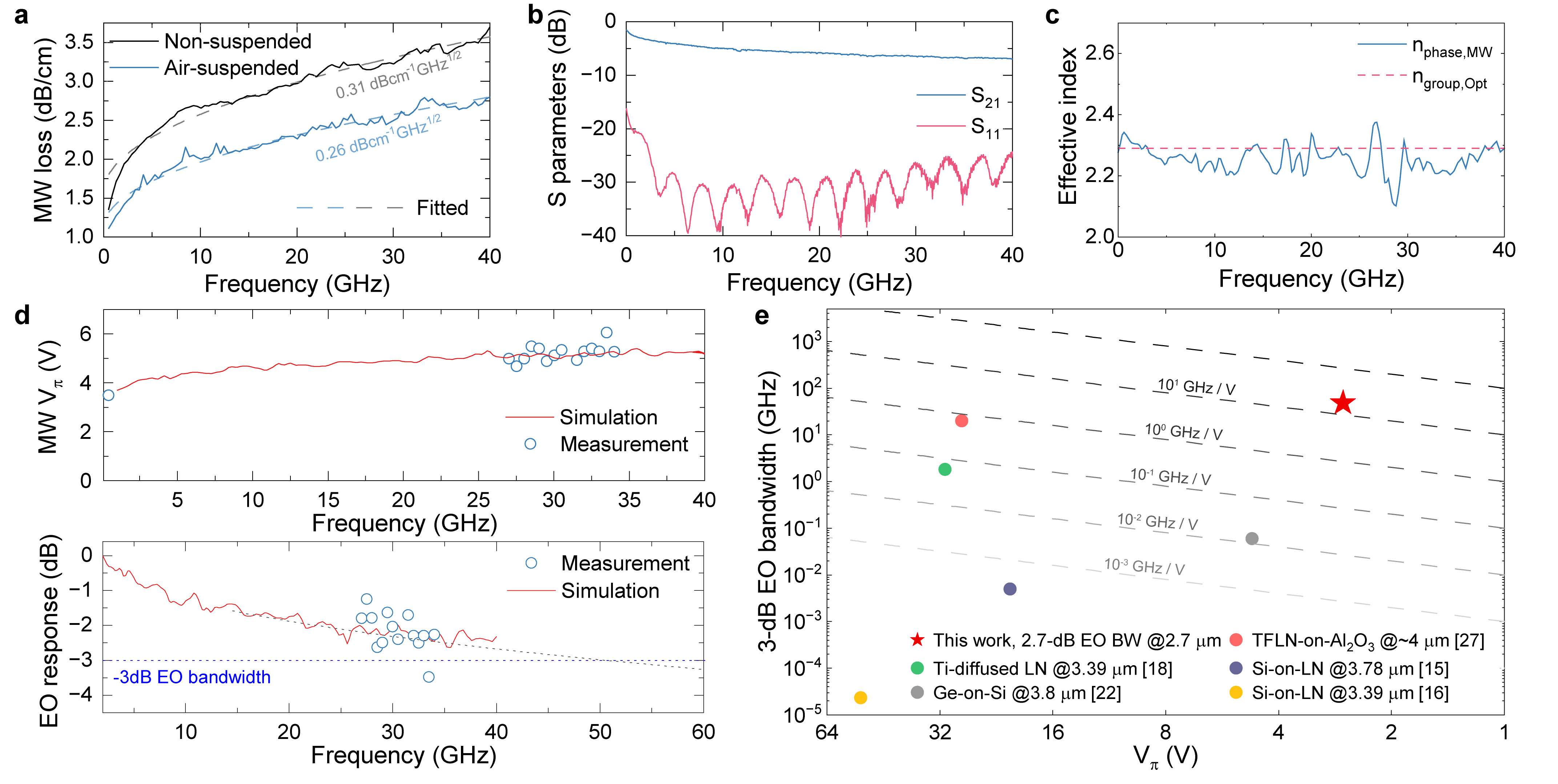}
\caption{\label{fig:figure3} \textbf{Characterization of the MIR modulator up to 40-GHz microwave frequencies.} (a) Measured microwave loss on the segmented co-planar electrodes before and after air suspension. Segmented travel-wave electrodes combined with air suspension lead to a reduced MW propagation loss of 2.75 dB/cm at 40 GHz and an ohmic-loss-limited slope of 0.26 dB/cm/$GHz^{0.5}$ , which is comparable to the loss slope of the best NIR LN modulators reported \cite{kharel2021}. (b) Measured electrical transmission $S_{21}$ and reflection $S_{11}$ spectrum. The measured reflection is below 25 dB from 2 - 40 GHz range indicating a well-matched impedance to 50 $\Omega$. (c) Measured MW phase index, which matches well with the optical group index based on the air-suspended TFLN waveguide (dashed line). (d) Microwave $V_\pi$ and the corresponding electro-optic response, referenced to the performance at 2 GHz.  Microwave $V_\pi$ are measured to be 5 V at 27 GHz and 5.26 V at 34 GHz for a 2-cm-long EO AM at the optical wavelength of 2.7 $\mu$m. The microwave $V_\pi$ is extracted via the optical spectrum from the AM under MW driving. The measurement results match with the simulation based on the measured S parameters in (a). Electro-optic response scales inversely with the square of the microwave $V_\pi$ and drops 2.7 dB at 40 GHz MW frequency which is the frequency limit of our signal generator while the extracted 3-dB EO BW is 50 GHz (dashed line). (e) Comparison of the modulator figure of merit, defined as the ratio between the 3-dB EO bandwidth (BW) and the near-DC $V_\pi$. The AM in our work demonstrates significantly higher BW/$V_\pi$ values of 17.4 GHz/V, as compared to the literature values (the dashed lines indicate constant BW/$V_\pi$ values). }
\end{figure}

Next, we demonstrate high performance of the suspended MIR modulator at microwave frequencies, including the EO response and the MW $V_\pi \cdot L$. We first measure the microwave attenuation loss, reflection loss and phase index of the segmented electrodes via a vector network analyzer in a frequency span from 10 MHz to 40 GHz, shown in Fig.~\ref{fig:figure3}a-c. The MW loss is extracted from $S_{21}$ and measured to be 2 dB/cm at 10 GHz and 2.75 dB/cm at 40 GHz, respectively. The fitted loss slope coefficient is  0.26 dB/cm/$GHz^{0.5}$ after air suspension (Fig.~\ref{fig:figure3}a). We discover a higher MW loss slope of 0.31 dB/cm/$GHz^{0.5}$  and an increased loss of 3.75 dB/cm at 40 GHz before air suspension of the waveguide (Fig.~\ref{fig:figure3}a), which suggests a potential loss channel induced by the bottom oxide layer. In conclusion, the record MW loss is a result of a thick Au metal thickness of 800 nm, optimization of the slow-wave T-shaped segment design, and removal of oxide layer. Secondly, Fig.3b also plots the reflection $S_{11}$ below -25 dB across 2 to 40 GHz range, indicating that the characteristic impedance of the transmission line is well matched to 50 $\Omega$ as designed. At last, Fig.~\ref{fig:figure3}c shows that the MW phase index, obtained by $n_{phase,MW}$ = ($c_{0}$×group delay)/ (electrode length), matches well with the optical group index ($n_{grp,optical}$) of 2.29. The achieved high performance electrodes contribute to a lower MW $V_\pi$ as the electrode length increases. The MW $V_\pi$ is extracted from the optical spectrum of the driven modulator output measured by a Fourier-transform infrared spectrometer. Figure~\ref{fig:figure3}d plots the measured MW $V_\pi$ of our 2-cm-long AM to be 4.5-6.5 V at 25-35 GHz frequency, which agrees well with our simulation based on the device parameters. The EO response can be simulated from the measured MW properties and scales inversely with the square of the microwave $V_\pi$, which matches well with the experimental data (Fig.~\ref{fig:figure3}d). The EO response drops 2.7 dB at 40 GHz MW frequency which is the frequency limit of our signal generator while the fitted 3-dB EO BW is 50 GHz (dashed line). To the best knowledge, this is the first time that the $V_\pi$ is ever measured at tens of GHz frequency range and presents the broadest EO bandwidth demonstrated so far in any MIR modulators. To note, another well-adopted figure-of-merit (FOM) is 3-dB EO bandwidth divided by near-DC $V_\pi$, which amortizes the electrode length dependence. In our work, this FOM is achieved at 17.4 GHz/V, orders of magnitude larger than the reported value (Fig.~\ref{fig:figure3}e). 

\begin{figure}[h]
\centering
\includegraphics[scale=1]{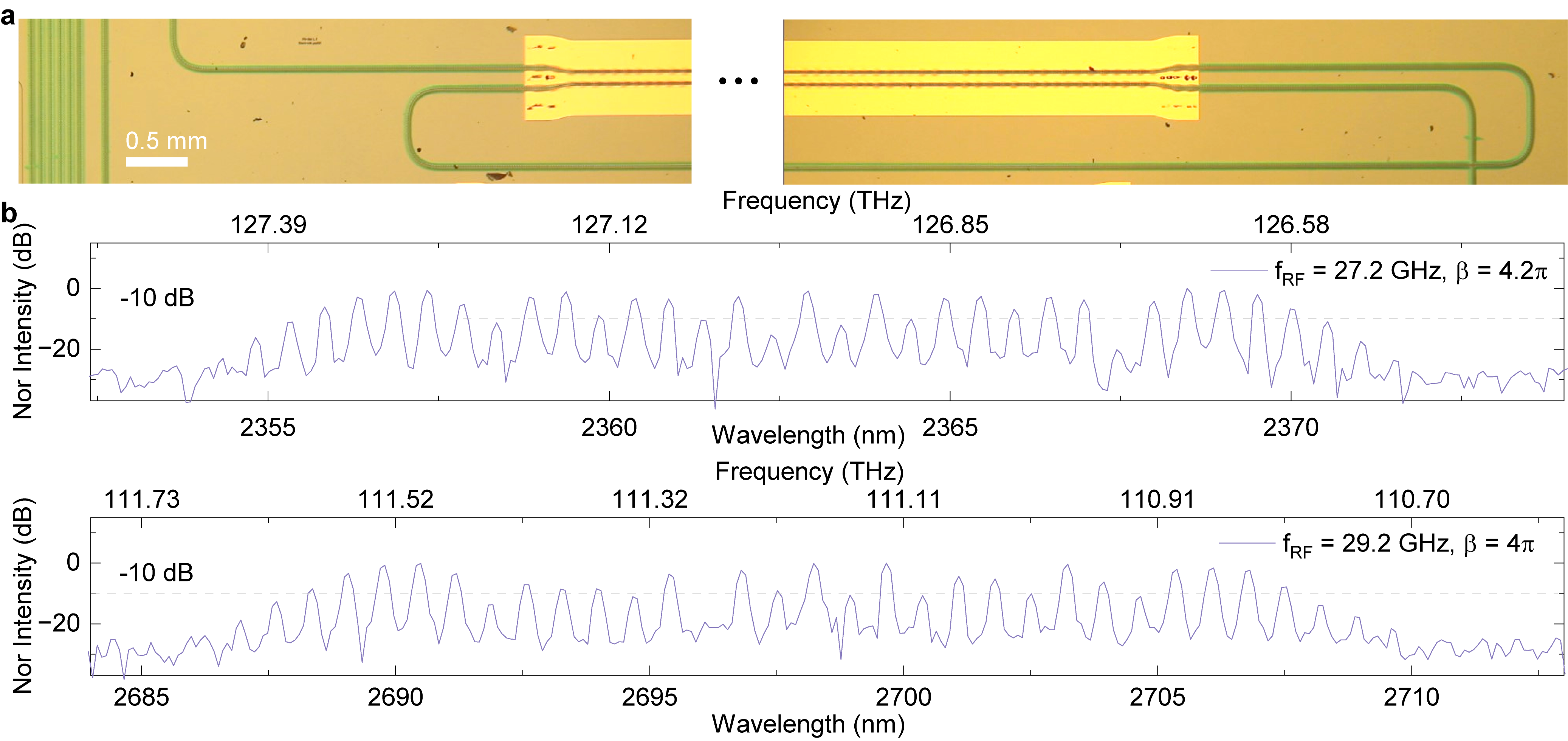}
\caption{\label{fig:figure4} \textbf{Frequency-agile MIR electro-optic frequency combs based on an integrated recycled phase modulator.} (a) Microscopic image of the suspended double passing phase modulator with the total modulation length of 4 cm. (b) Broadband EO frequency comb generation using the same PM device at two different MIR wavelengths of 2.36 $\mu$m and 2.7 $\mu$m. Total modulation indexes of 4.2 $\pi$  and 4 $\pi$ at MW driving frequencies of 27.2 and 29.2 GHz were measured at 2.36 $\mu$m and 2.7 $\mu$m pump wavelength, respectively. This corresponds to a 10-dB optical bandwidth of 0.8 THz.}
\end{figure}

In addition to the amplitude modulation, we demonstrate a suspended 2-cm-long TFLN PM and generate frequency agile EO frequency combs in the MIR.  Figure~\ref{fig:figure4}a shows the optical image of the fabricated PM where we adopt a double-pass photonic structure where the waveguide is rerouted into the same transmission line after a waveguide crossing. Therefore, the total active modulation length is achieved at 4 cm. Further leveraging the low MW $V_\pi$ and large EO bandwidth of our platform, we generate an integrated MIR EO frequency comb with a 10-dB optical bandwidth of 0.8 THz (20 nm) and a total comb line number of 33 at a center wavelength of 2.7 $\mu$m by driving the PM with a single-frequency MW signal at 29.2 GHz (Fig.~\ref{fig:figure4}b). In addition, by driving the same device, we achieve an EO comb spectrum at 2.36 $\mu$m with a 27.2-GHz line spacing and a similar 10-dB span of 0.8 THz (15 nm), which shows the frequency agility of the non-resonant waveguide-based MIR EO comb generator in both center wavelengths and comb line spacings. The total modulation index ($\beta$) we achieve are 4.2 $\pi$  and 4 $\pi$ at 27.2 GHz and 29.2 GHz, respectively. This is the first demonstration of integrated EO comb generation in the MIR with tens of GHz repetition rates, without any need of parametric conversion. Complementary to the semiconductor based combs \cite{heckelmann2023,kazakov2025,zeng2025} and Kerr microcombs in the MIR \cite{yu2016,yu2017}, the EO comb sources offer unique advantages including tunable spectral resolution, wavelength multiplexing and robust mode-locking operation for sensing and communication applications \cite{yan2017}.

\begin{figure}[h]
\centering
\includegraphics[scale=1]{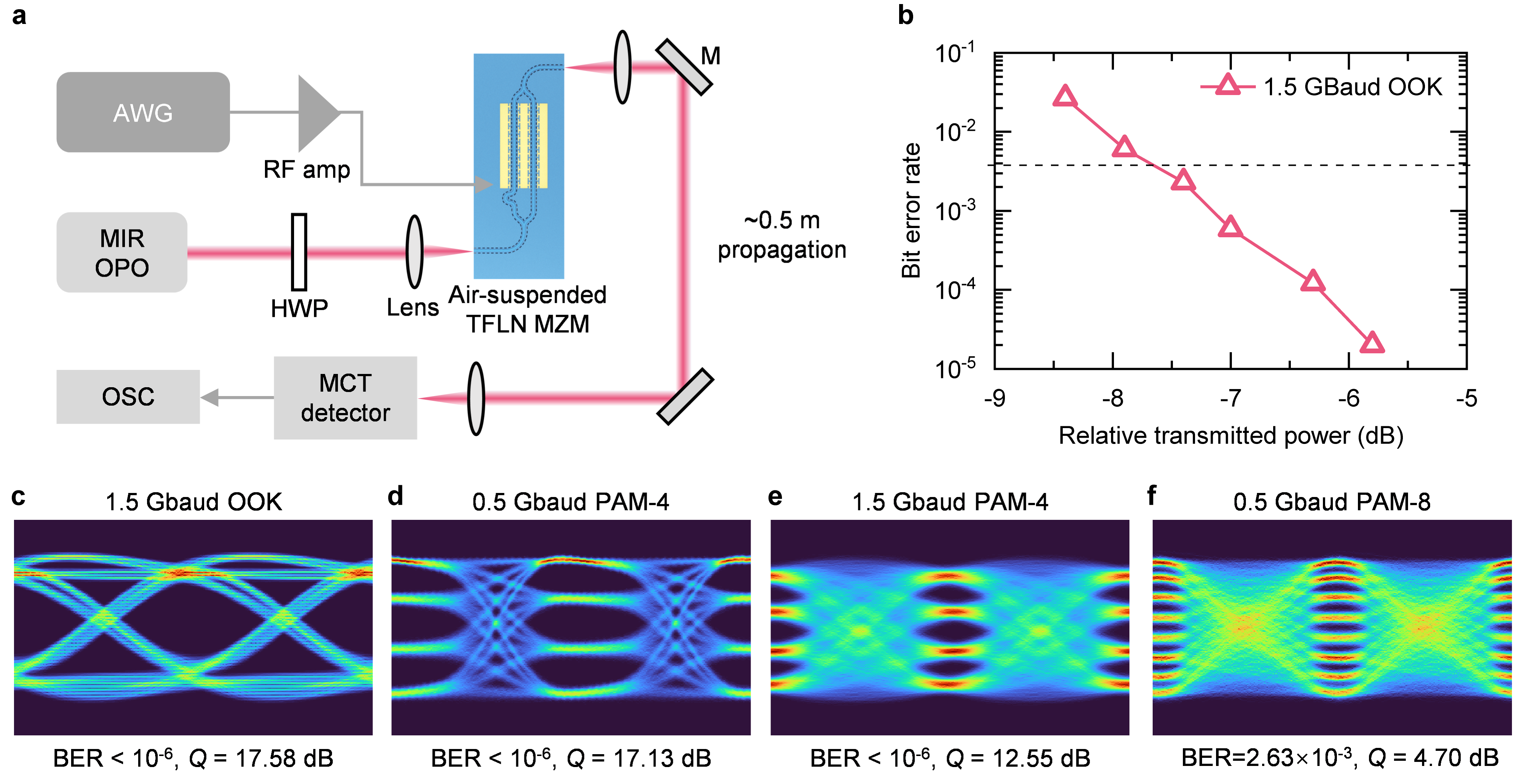}
\caption{\label{fig:figure5} \textbf{Experimental MIR communication link using an air-suspended TFLN intensity modulator.} (a) Experimental setup for measuring the communication link at 2.7 $\mu$m wavelength. OPO: optical parametric oscillator, HWP: half-wave plate, AWG: arbitrary waveform generator, MCT detector: mercury cadmium telluride detector, DSO: digital storage oscilloscope. (b) Measured BER curve versus transmitted optical power for 1.5-Gbaud on-off keying (OOK) signal. (c-f) Experimental results of eye diagram, BER values, and Q-factors of (c) 1.5-Gbaud OOK, (d) 0.5-Gbaud Pulse Amplitude Modulation (PAM)-4, (e) 1.5-Gbaud PAM-4, (f) 0.5-Gbaud PAM-8. The achievable data rate is limited by the 3-dB bandwidth of our MIR detector (1 GHz). }
\end{figure}

Finally, we evaluate a free space communication link based on the suspended TFLN AM as a proof of principle demonstration. We experimentally demonstrate an intensity modulation/direct detection (IM/DD) MIR link at 2.7 $\mu$m. Figure~\ref{fig:figure5}a illustrates the experimental setup (see Methods). The modulated MIR output propagates 0.5 m in the free space and is detected by the MCT detector with a 3-dB bandwidth of $>$ 1 GHz. We first measure the BER performance of OOK signals at various input power levels to the AM with a baud rate of 1.5 Gbaud, as shown in Fig.~\ref{fig:figure5}b. The reference case (0 dB) corresponds to the maximum transmitted power. In this setup, the system performance is primarily limited by the detector bandwidth and total loss resulting in a BER decrease from $10^{-2}$ to $10^{-5}$ within a 2.5-dB power range. Additionally, we demonstrate higher-order intensity modulation formats at a maximum transmitted power, including 4-level and 8-level pulse amplitude modulation (PAM-4 and PAM-8). Eye diagrams, BER values, and Q-factors for different modulation formats and baud rates are presented in Fig.~\ref{fig:figure5}(c--f) for comparison. All configurations achieved BER values below the 7$\%$ hard-decision forward error correction (HD-FEC) threshold. The phase modulator demonstrated on the same platform further highlights its capability for phase-modulated formats and more complex data modulation.

In conclusion, the suspended TFLN EO devices demonstrated here are prototypical building blocks for a new generation of MIR integrated photonic architectures, expanding functionality beyond amplitude modulation to include phase modulation and frequency-comb generation. We achieved record-low $V_\pi$  at both near-DC and previously unexplored tens-of-GHz microwave frequencies range, a broad EO bandwidth exceeding 40 GHz, and an ultra-wide optical window covering 1.55 µm and 2.4 to 3.6 µm in the MIR.  This work marks the first time that an MIR EO modulator exceeds the centimeter-scale interaction length while achieving a modulation index exceeding 4$\pi$ and a FOM of 17.4 GHz/V for the EO bandwidth per drive voltage, more than an order of magnitude improvement over other MIR platforms (Table 1). Leveraging these unique characteristics, we demonstrated the first, to our knowledge, direct EO frequency comb generation in the MIR with a 0.8-THz span and over 33 comb lines. These results establish TFLN as a compelling monolithic platform for scalable MIR EO systems. The suspended platform could be extended to operate beyond 3.6 µm to explore the full LN transparency window (up to 5 µm) by increasing the electrode gap and applying the same co-design principle. The optical loss can be improved via further optimizing the facet design and fabrication process with thermal and chemical treatment \cite{zhu2024,shamsansari2022,he2019}. A straightforward yet crucial next step will be to integrate in-phase/quadrature (IQ) modulator configurations for coherent MIR communication and vector signal processing. We highlight the versatility of the TFLN platform where periodically poled TFLN waveguides would further be integrated to down-convert the high speed MIR signal to the NIR as the receiver of an FSOC link. Further integration of the AM and PM on the same chip will also enable MIR pulse generation and on-chip waveform synthesis \cite{yu2022}. The demonstrated comb bandwidth of 0.8 THz would lead to a 1.7-ps pulse train at 2.7 µm at a 30-GHz repetition rate via EO modelocking. On-chip EO modulation offers a path to stabilized MIR comb sources and synchronized pulse trains \cite{yu2022, alirezalaser}, bridging the gap between semiconductor lasers and electro-optic frequency synthesis. With continued advancement of QCLs and chip-based parametric oscillators \cite{hwang2023,chen2023}, we envision a hybrid passive and active nonlinear integrated system where frequency conversion, EO modulation and comb generation could coexist. Such systems would not only enable spectrally tailored dual-comb spectroscopy for molecular sensing and metrology \cite{Shams-Ansari2022} but also provide multi-wavelength coherent sources for free-space optical communication \cite{zou2022,pang2022,su2023} and astronomical spectrograph calibration \cite{sekhar2024,metcalf2019}. The ability to combine low-loss EO modulation, broadband MIR transparency, and scalable chip-level integration marks an important step toward fully reconfigurable, multifunctional MIR photonic systems.


\clearpage
\begin{table*}[h]
\centering
\caption{Comparison of mid-infrared electro-optic modulators.}
\footnotesize
{\fontsize{7}{9.5}\selectfont 
\renewcommand{\arraystretch}{1.2}
\setlength{\tabcolsep}{1pt}
\begin{adjustbox}{width=\textwidth}
\begin{tabular}{l l l c c c c c c c c c c}
\toprule
\makecell{\textbf{Wavelength} \\[2pt]\textbf{(\boldmath{$\mu$m})}} &
\makecell{\textbf{Platform}\\[2pt]\textbf{design}} &
\makecell{\textbf{Scheme}} &
\makecell{\textbf{Function}} &
\makecell{\textbf{L}\\\textbf{(cm)}} &
\makecell{\textbf{$V_{\pi,DC}$}\textbf{(V)}} &
\makecell{\textbf{$V_{\pi,MW}$}\textbf{(V)}} &
\makecell{\textbf{Optical}\\[2pt]\textbf{loss}\\\textbf{(dB/cm)}} &
\makecell{\textbf{MW loss}\\[2pt]\textbf{(dB/cm)}} &
\makecell{\textbf{EO}\\[2pt]\textbf{BW}} &
\makecell{\textbf{EO BW}\\[2pt]\textbf{per $V_{\pi,DC}$}\\[2pt]\textbf{(GHz/V)}}&
\makecell{\textbf{Application}} &
\textbf{Ref.}\\
\midrule

\makecell{2.4\\--3.6} &
\makecell{Air-suspended\\[2pt]TFLN, MZI} &
Pockels &
\makecell{AM,\\[2pt]PM,\\[2pt]combs} &
2 &
\begin{tabular}{@{}c@{}}2.3-4.2\end{tabular} &
\begin{tabular}{@{}c@{}}4.5-6.5$^{b}$ \end{tabular} &
\begin{tabular}{@{}c@{}}2.2-2.8\end{tabular} &
\begin{tabular}{@{}c@{}}2.75 \\[2pt] @40GHz\end{tabular} &
{\begin{tabular}{@{}c@{}}\makecell{40~GHz$^{c}$\\[2pt]50~GHz$^{d}$}\end{tabular}} &
17.4 &
\makecell{EO comb 0.8-THz span\\[2pt]Mod index of $4\pi$\\[2pt]OOK@1.5~Gbaud\\[2pt] PAM8@1.5~Gbaud} &
\makecell{\textbf{This}\\[2pt] \textbf{work}} \\
\midrule

\makecell{2.5} & \makecell{TiO$_2$-on-LN,\\[2pt]WG}& Pockels & AM & NA & NA & NA & NA & NA & NA & NA & NA & \cite{jin2019} \\
\makecell{2.6} & \makecell{BTO-on-oxide,\\[2pt]WG} & Pockels & AM & 0.2 & 55 & NA & 2.3 & NA & NA & NA &NA & \cite{jin2019BTO} \\
\makecell{3.39} & \makecell{Si-on-LN,\\[2pt]MZI} & Pockels & AM & 0.5 & 52 & NA & \makecell{2.5} & NA & $>$23~kHz$^{a}$ & 4.4$\times10^{-7}$ &NA & \cite{chiles2014} \\
\makecell{3.39} & \makecell{Ti-diffused LN,\\ [2pt]MZI} & Pockels & AM & 0.8 & 31 & NA & NA & NA & 1.8~GHz & 0.058 & NA & \cite{becker1985} \\
\makecell{3.78} & \makecell{Si-on-LN,\\[2pt]MZI} & Pockels & AM & 0.6 & 20.8 & NA & \makecell{4.5} & NA & $>$5~MHz$^{a}$ & 0.00024 &NA & \cite{xu2023} \\
\makecell{3.8} & \makecell{Si-on-SiO$_2$, \\MZI} & TO & AM & NA & NA & NA & 3.5 & NA & 23.8~kHz & NA & NA & \cite{nedeljkovic2014} \\
\makecell{3.85\\--4.1} & \makecell{BP-Si-on-SiO$_2$,\\[2pt]WG} & EA & AM & NA & NA & NA & NA & NA & $>$60~kHz & NA & NA & \cite{huang2021} \\
\makecell{3.8} & \makecell{Ge-on-Si,\\[2pt]MZI/EAM} & Cl/CA & AM & 0.1 & 4.7 & NA & NA & NA & $>$60~MHz & 0.013 & OOK@60~MHz & \cite{li2019} \\
\makecell{3.72\\--3.8} & \makecell{Si-on-oxide,\\[2pt]MZI/EAM} & Cl/CA & AM & NA & NA & NA & 11.5 & NA & NA & NA & OOK@125~Mbit/s & \cite{nedeljkovic2019} \\
\makecell{3.95\\--4.3} & \makecell{TFLN-on-Al$_2$O$_3$,\\[2pt]MZI} & Pockels & AM & 0.8 & 28 & NA & 2.2 & NA & 20~GHz & 0.71 & OOK@10~Gbit/s & \cite{didier2025} \\
\makecell{3.3\\--3.8} & \makecell{Air-clad TFLN,\\[2pt]cavity} & Pockels & AM & 0.15 & 51 & NA & 4 & NA & NA & NA & NA & \cite{hwang2025} \\

\bottomrule
\end{tabular}
\end{adjustbox}
\caption*{\footnotesize 
ER: extinction ratio; EO: electro-optic; TO: thermo-optic; EA: electro-absorption;
BP: black phosphorus; EAM: electro-absorption modulator; 
CI: free carrier injection; CA: free carrier absorption. Mod: modulation\\[2pt]
$^{a)}$ The measured bandwidths are limited by the PD speed.\\
$^{b)}$ $V_{\pi}$ measured at 2.7 $\mu$m wavelength and 25-35 GHz MW frequency.\\
$^{c)}$ 2.7~dB EO BW. Limited by the VNA performance.\\
$^{d)}$ Extrapolated.

Note:  During the preparation of this manuscript, a related work was posted online \cite{didier2025} which performance are cited in this table.}
}
\end{table*}

\section{Methods}
\subsection{Device Fabrication}

To support guided modes in the MIR spectrum, we patterned waveguides using electron beam (e-beam) lithography with a thick hydrogen silsesquioxane (HSQ) resist and dry etched 500~nm of lithium niobate (LN) using reactive ion etching (RIE). Air holes were patterned near the LN waveguides using direct laser writing and etched with RIE through a photoresist mask. Subsequently, 800~nm-thick electrodes were patterned with e-beam lithography, followed by metal deposition using a thermal evaporator and a lift-off process. The underlying SiO$_2$ layer was then etched using a buffered oxide etch (BOE) solution.


\subsection{Wavelength dependent near-DC $V_\pi$}

\begin{equation}
\Delta \beta_{\mathrm{TE}} 
= \frac{2 n_e \, \Delta n_e}{2 \beta_{\mathrm{TE}}} k_0^{2} 
= -\frac{n_e^{4} r_{33} {\Gamma_{\mathrm{mo}}}}{2 \beta_{\mathrm{TE}}} k_0^{2}
\end{equation}

\begin{equation}
V_{\pi} \cdot L = \frac{\pi}{\Delta \beta_{\mathrm{TE}}} \times 100~(\mathrm{V \cdot cm})
\end{equation}
\noindent
\textbf{Definitions:}
\begin{itemize}
    \item $\Delta \beta_{\mathrm{TE}}$: total phase change per meter per volt for TE guided mode.
    \item $k_0$: wavevector in vacuum.
    \item $\beta_{\mathrm{TE}}$: propagation constant of the guided mode.
    \item $n_e$: extraordinary refractive index of LiNbO$_3$ (LN).
    \item $r_{33}$: electro-optic (EO) coefficient.
    \item $\Gamma_{\mathrm{mo}}$: mode overlap factor.
\end{itemize}

Dispersive DC $V_{\pi}$ is studied using COMSOL finite element method numerical simulation using equation (1) and (2). Extended Data Figure~1a shows comparison between measurement and simulation results. Trenching effect at the bottom of waveguide sidewall is included in the simulation. Measured DC $V_{\pi}$ values matches well to the simulation in all wavelength ranges. We note that a higher $V_{\pi}$ is expected at a larger optical wavelength due to the smaller wave-vector as well as a smaller mode overlap between optical and microwave fields (Extended Data Fig. 1b).

\begin{figure}[h]
\centering
\includegraphics[scale=1]{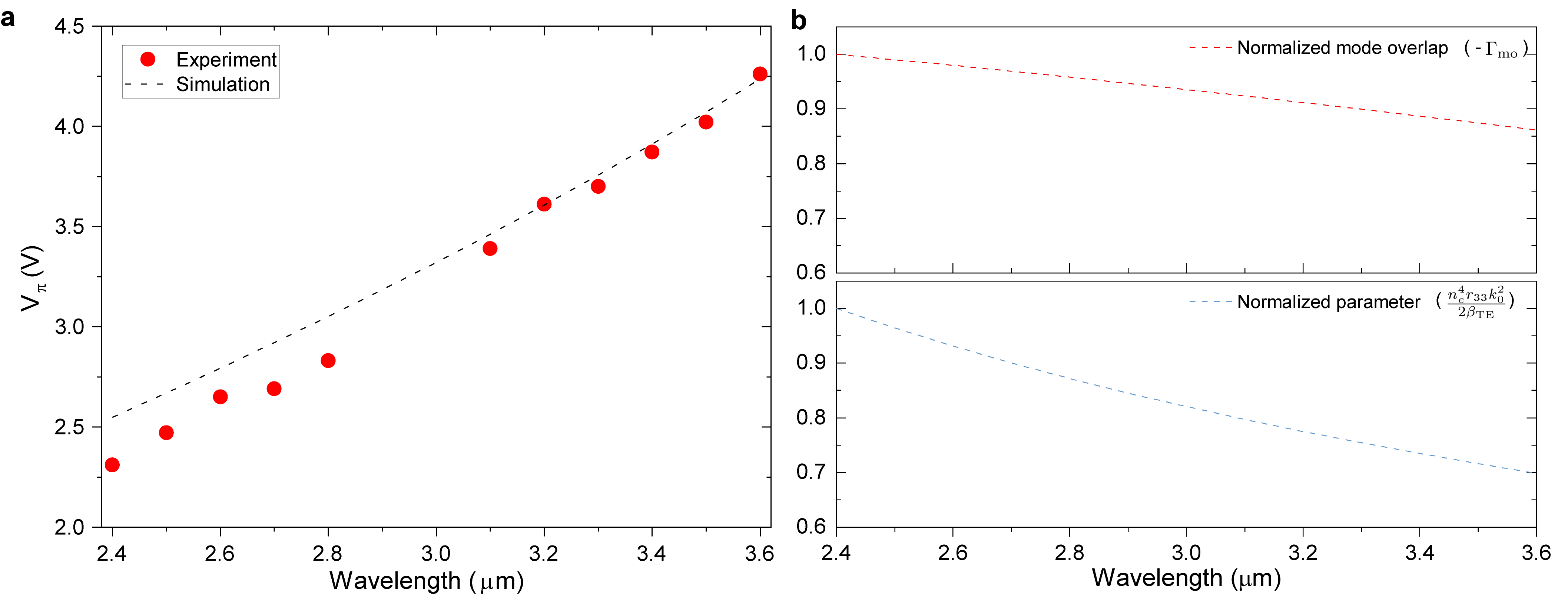}
\captionsetup{labelformat=empty} %
\caption{\label{fig:extended_figure2} \textbf{Extended Data Fig. 1}: Analysis for wavelength-dependent DC $V_{\pi}$.} (a) DC $V_{\pi}$ measurement and simulation for the wavelength range of 2.4--3.6~$\mu$m. (b) Simulated mode overlap factor and other parameters used for $\Delta \beta_{\mathrm{TE}}$ calculation.
\end{figure}

\subsection{Extinction ratio analysis}
The extinction ratio is measured using a chopper, a MIR photodetector (PVI-4TE-8), and a lock-in amplifier as the AWG sweeps the applied voltage on the AM. After the sinusoidal fitting, the extinction ratio is 7.1~dB based on the maximum and minimum transmission values (Extended Data Fig.~2). The extinction ratio is primarily limited by the higher-order mode coupling to the slab at the second Y-splitter/merger, which can be improved by using a single-mode waveguide width of 0.9~$\mu$m instead of 2~$\mu$m in our case.
\begin{figure}[h]
\centering
\includegraphics[scale=1]{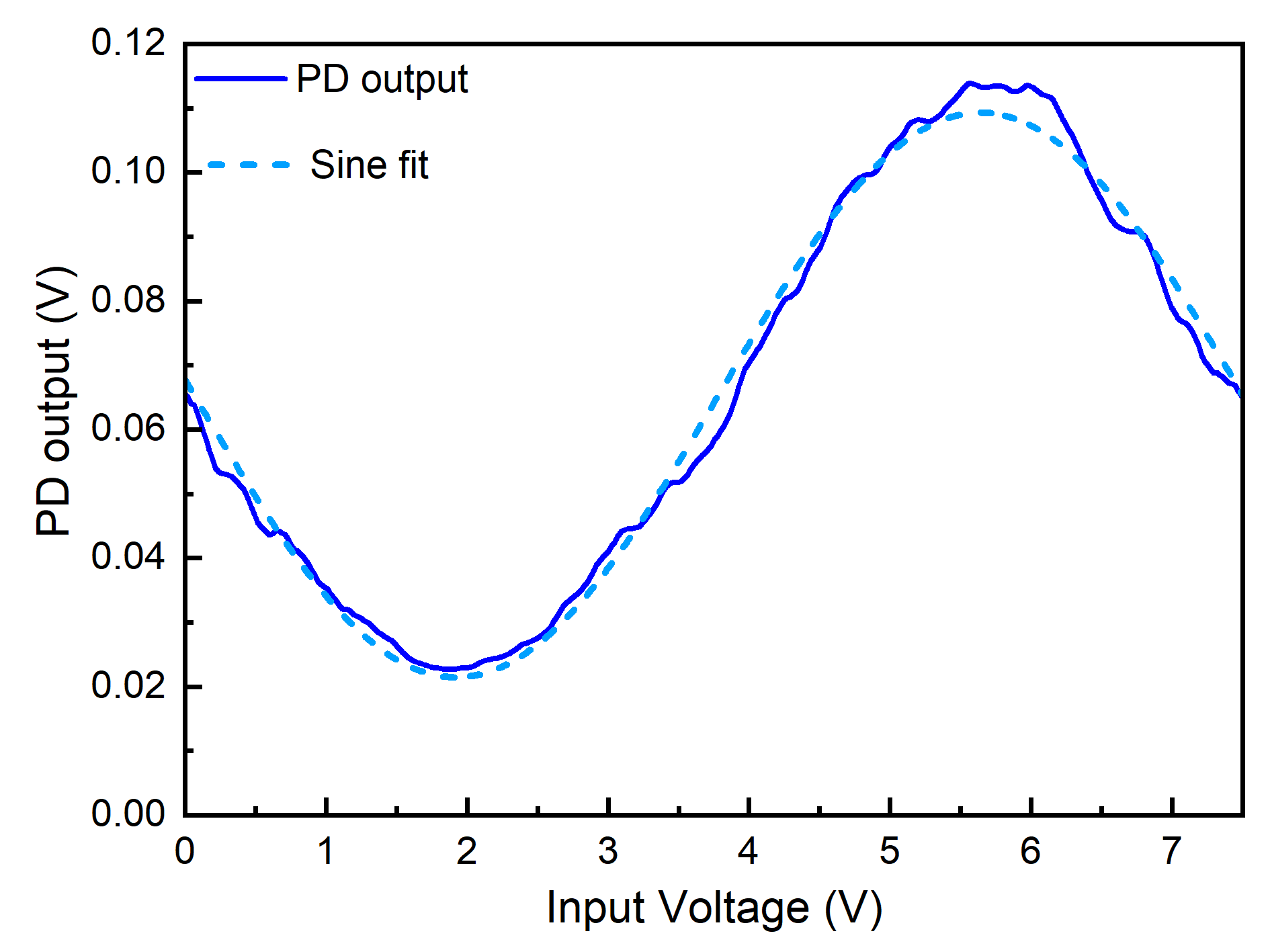}
\captionsetup{labelformat=empty} %
\caption{\label{fig:extended_figure2} \textbf{Extended Data Fig. 2}: 
Extinction measurement of the air-suspended amplitude modulator at 2.7~$\mu$m wavelength.}
\end{figure}

\subsection{Communication Setup}

A TOPTICA continuous-wave (CW) optical parametric oscillator (OPO) is utilized in our system as a mid-infrared (MIR) source. Using a \SI{1064}{\nano\meter} CW pump laser, this OPO can generate tunable $\sim$\SI{1}{\watt} MIR power within the range of 2.5--4~\si{\micro\meter} with a laser linewidth of approximately \SI{500}{\kilo\hertz}. The free-space output of the OPO is coupled into the integrated Mach–Zehnder amplitude modulator (AM) using a MIR collimating lens. 

An arbitrary waveform generator (AWG) with a sampling rate of \SI{20}{GSa/s}, followed by an electrical amplifier with \SI{26}{\decibel} gain, is used to apply various intensity modulation signals to the AM. The electrical signal is delivered to the device electrode through a \SI{67}{\giga\hertz}-bandwidth RF probe. After free-space propagation of approximately \SI{0.5}{\meter}, the output of the modulator is focused onto a MIR mercury cadmium telluride (MCT) detector for detection. This MCT detector has a bandwidth of about \SI{1}{\giga\hertz}. 

The detector output is recorded by a digital storage oscilloscope (DSO) for data recovery. Specifically, offline digital signal processing (DSP) is applied to compensate for system degradation in multi-level pulse amplitude modulation (PAM) signals. A 20-tap adaptive filter based on recursive least-squares (RLS) equalization with a training sequence is used to improve signal quality. No additional DSP equalization is applied to the on–off keying (OOK) signal in this measurement.

\subsection{Insertion Loss Analysis}

We use two free-space objective lenses (Thorlabs, C036TME-E) to couple the MIR light into and out of the devices over a wavelength range of 2.4--3.6~$\mu$m, with an off-chip optical power ranging from 60 to 700~mW at the chip facet. Extended Data Table~1 summarizes the optical losses measured from both the suspended device used in the main manuscript and non-suspended reference devices (air clad) of identical dimensions. The passive and active propagation losses were extracted from waveguides, amplitude modulators and recycled phase modulators with different waveguide lengths. As shown in the Extended Data Table 1, the optical loss at 2.4~$\mu$m remains comparable before and after suspension, while the suspended devices exhibit noticeably lower loss at 2.7, 2.8 and 3.6 $\mu$m compared to the non-suspended counterpart. This result agrees well with the spectral absorption feature of SiO$_2$ in the mid-infrared where the absorption has a peak around 2.7-2.8 $\mu$m and increases starting at 3.1 $\mu$m. Even though our device is designed and measured up to 3.6$\mu$m, the suspended TFLN approach is expected to continuously reduce the propagation loss beyond 3.6$\mu$m up to the end of the transmission window of LN at 4.5-5 $\mu$m.

However, the active waveguides, which include nearby metal electrodes for electro-optic modulation, show higher loss than the passive sections. The measured active propagation losses at 2.4~$\mu$m and 3.6~$\mu$m are 6.4~dB/cm and 10.5~dB/cm, respectively. The exact cause of this additional attenuation is still under investigation as the simulated metal induced loss is 0.2 and 4.8 dB/cm respectively. One of the possible reason is the high energy electron injection during the ebeam lithography of the metal electrodes nearby the waveguide which induces higher free carriers loss at higher optical wavelength, which could be improved in the future via controlled annealing condition and photolithography.

\begin{table}[h]
\centering
\centering
\captionsetup{labelformat=empty} 
\caption{\textbf{Extended Data Table 1.} Measured propagation loss of the fabricated devices.}
\begin{tabular}{lccc}
\toprule
\textbf{} & \textbf{Wavelength ($\mu$m)} &
\makecell{\textbf{Propagation loss}\\\textbf{passive (dB/cm)}} \\
\midrule
\textbf{Suspended} & & & \\
\midrule
 & 2.4 & 2.2 \\
 & 2.7 & 1.2 \\
 & 2.8 & 4.6 \\
 & 3.6 & 2.8 \\
\midrule
\textbf{Non-suspended} & & & \\
\midrule
 & 2.4 & 2.4 \\
 & 2.7 & 7.2 \\
 & 2.8 & 11.4 \\
 & 3.6 & 4.3 \\
\bottomrule
\end{tabular}
\end{table}

\section{Data availability}

The data that support the findings of this study are available from the corresponding author upon reasonable request.

\section{Acknowledgements}

The authors thank Alexander Gaeta and Yun Zhao for providing the mid-infrared photodetector. This work is supported by the Optica Foundation, the Chan Zuckerberg Initiative Foundation(Dynamic Imaging, 2023-321175), and the DARPA Young Faculty Award (D23AP00252-02). M.Y. and Y.Y. are supported by the U.S. Department of Energy, Office of Science, Basic Energy Sciences, Materials Sciences and Engineering Division under Contract No. DE-AC02-05CH11231 within the Quantum Coherent Systems Program KCAS26. AW is supported by the Office of Navel Research through MURI Award(N00014-20-1-2558) and Airbus Institute for Engineering Research. Device fabrication was performed at the John O’Brien Nanofabrication Laboratory at Universtity of Southern California. The views, opinions and/or findings expressed are those of the authors and should not be interpreted as representing the official views or policies of the Department of Defense or the U.S. Government. 

\section{Author contributions}

M.Y. conceived the conceptual idea. C.L. designed the chip and fabricated the devices. C.L., X.R. and X.S. designed the experiment and carried out the measurements. C.L., X.R., X.S. and M.Y. analyzed the data with the help of W.L., Z.J., Y.Y., H.Z., Y.Z., S.O., R.K., A.T.H., M.T., A.E.W., and Z.C.. C.L., X.R., X.S. and M.Y. drafted and revised the manuscript with contribution from all authors. M.Y., Z.C. and A.E.W. supervised the project.

\section{Competing interests}

C. L., Z.C. and M.Y. are involved in developing lithium niobate technologies at Opticore Inc..

\clearpage
\bibliographystyle{unsrt}
\bibliography{Reference} 

\end{document}